\documentclass[fleqn,usenatbib]{mnras}

\usepackage{newtxtext,newtxmath}

\usepackage[T1]{fontenc}
\usepackage{ae,aecompl}
\usepackage{ulem}
\usepackage{setspace}
\usepackage{graphicx}	
\usepackage{amsmath}	

\usepackage{arydshln}

\usepackage[dvipsnames]{xcolor}

\newcommand{\orcid}[1]{\href{https://orcid.org/#1}{\includegraphics[width=0.7em]{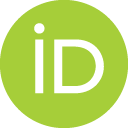}}}

\def\mthod{$\mathrm{MTHOD}\,$}
\def\mpcoh{\,h^{-1}{\rm Mpc}}
\def\msolaroh{\,h^{-1}M_\odot}

\def\gtsim{\mathrel{\lower0.6ex\hbox{$\buildrel {\textstyle >}
 \over {\scriptstyle \sim}$}}}
\def\ltsim{\mathrel{\lower0.6ex\hbox{$\buildrel {\textstyle <}
 \over {\scriptstyle \sim}$}}}

\title[QSO physics with eBOSS]{Quasars at intermediate redshift are not special; but they are often satellites}

\author[Alam et al.]
{Shadab Alam \orcid{0000-0002-3757-6359},$^{1}$\thanks{E-mail: salam@roe.ac.uk}
Nicholas P. Ross,$^{1}$
Sarah Eftekharzadeh,$^{2}$
John A. Peacock,$^{1}$
\and Johan Comparat,$^{3}$
Adam D. Myers,$^{4}$
and Ashley J. Ross$^{5}$
\\
$^{1}$Institute for Astronomy, University of Edinburgh, Royal Observatory, Blackford Hill, Edinburgh, EH9 3HJ , UK\\
$^{2}$Department of Physics and Astronomy, University of Utah, 115 S 1400 E, Salt Lake City, UT 84112\\
$^{3}$Max-Planck-Institut f\"{u}r extraterrestrische Physik (MPE), 
Giessenbachstrasse 1, D-85748 Garching bei M\"unchen, Germany\\
$^{4}$Department of Physics and Astronomy, University of Wyoming, Laramie, WY 82071, USA\\
$^{5}$Center for Cosmology and Astro-Particle Physics, Ohio State University, Columbus, Ohio, USA\vspace*{-4pt} \\
}

\pubyear{2020}

\begin{document}
\label{firstpage}
\pagerange{\pageref{firstpage}--\pageref{lastpage}}
\maketitle

\begin{abstract}
Understanding the links between the activity of supermassive black holes (SMBH) at the centres of galaxies and their host dark matter haloes is a key question in modern astrophysics.
The final data release of the SDSS-IV eBOSS provides the largest contemporary spectroscopic sample of galaxies and QSOs. Using this sample and covering  the redshift interval $z=0.7-1.1$, we have measured the clustering properties of the eBOSS QSOs, Emission Line Galaxies (ELGs) and Luminous Red Galaxies (LRGs). 
We have also measured the fraction of QSOs as a function of the overdensity defined by the galaxy population. Using these measurements, we investigate how QSOs populate and sample the galaxy population, and how the host dark-matter haloes of QSOs sample the underlying halo distribution. We find that the probability of a galaxy hosting a QSO is independent of the host dark matter halo mass of the galaxy. We also find that about 60\% of eBOSS QSOs are hosted by LRGs and about 20-40\% of QSOs are hosted by satellite galaxies. We find a slight preference for QSOs to populate satellite galaxies over central galaxies. This is connected to the host halo mass distribution of different types of galaxies. Based on our analysis, QSOs should be hosted by a very broad distribution of haloes, and their occurrence should be modulated only by the efficiency of galaxy formation processes.
\end{abstract}



\section{Introduction}
Two  outstanding questions in extragalactic astrophysics are the manner in which galaxies sample the dark matter (DM) halo mass
function, and how active galactic nuclei (AGN) sample the galaxy population. These questions are central because it is now believed that the energy and kinematics associated with AGN are crucial in understanding how
galaxies form and regulate their star formation \citep[see reviews by
e.g.][]{Fabian2012, KormendyHo2013, MadauDickinson2014,
KingPounds2015, SomervilleDave2015, Xue2017, Padovani2017}.

Although they are rare, QSOs (generally defined as luminous AGN with bolometric
luminosities $L_{\rm bol}$ above $\sim$10$^{38}\,$W) have
become key tracers of the large-scale structure (LSS) of the Universe
\citep[e.g.][]{Coil2004, Outram2004, Croom2005, Myers2007a, Shen2007, daAngela2008, Ross2009, He2018, neveux20a, hou20a,2020MNRAS.tmp.2072O}. However, the detailed link between central black holes and their DM haloes 
is still poorly understood.  

In the case of a Gaussian random field, the two-point correlation function 
statistic provides a full characterisation of large-scale structure
\citep[see e.g.][]{Peebles1980book, Bardeen1986, Wang2013}. A given LSS tracer will display biased clustering, with an amplitude that increases for objects associated with rare massive haloes. Thus, by measuring the 2PCF of QSOs and assuming an underlying form of the dark matter halo
mass function, it is possible to associate the QSOs with a given DM halo mass. The bias parameter, $b$, is typically determined on linear scales $\sim5 - 30 \mpcoh$, and this
measurement and comparison has been carried out for a range of QSO redshifts, luminosities and colours. 

Results from the last 10-20 years have traditionally placed QSOs 
in a mean DM halo mass of a few $\times10^{12} M_{\odot}$
\citep[e.g.][]{Croom2005, Coil2007, Myers2007a, daAngela2008, Ross2009}. 
However, one critical drawback of these measurements is that often
only a single `typical' halo mass is reported;
given the very large range of black hole masses in QSOs ($M_{\rm BH}$ = $10^6
- 10^{10} \msolaroh$), this single effective halo mass is relatively uninformative.  
Ideally, one would like to understand the full distribution of halo masses associated with QSOs, and then relate this to
the underlying dark matter halo mass functions, and indeed to the way in which galaxies in general populate haloes. Are QSOs a subset of 
the full galaxy population? Or is luminous AGN activity found 
preferentially in one type of galaxy? Or in a particular environment?

An exception to the `single halo mass' approach is \citet{White2012} who
assume that QSOs reside in haloes with a lognormal distribution of
masses, centred on a characteristic mass that scales with
luminosity. This model results in a range of masses for luminous QSOs at
redshift $z\approx 2.4$ from $0.8 - 4 \times 10^{12} \msolaroh$, 
with a central value of $2 \times 10^{12} \msolaroh$. \cite{2011ApJ...726...83M} and \cite{Kru12} also studied the cross-correlation of ROSAT AGN with SDSS galaxies and constrain the host halo mass distribution of the ROSAT AGN. They predict that the satellite fraction of AGN reduces with host halo mass in contrast to luminous galaxies.

Several groups and authors have used the 2PCF to infer 
the halo occupation distribution (HOD) of the QSO population, and the `satellite fraction'. 
The HOD provides a complete description of the relation between QSOs and dark matter at the 
level of individual virialized haloes \citep{Berlind2002, Kravtsov2004, Zheng2005, Chatterjee2012, Richardson2012}. 
The QSO HOD is defined by $P(N|M_{h})$, the conditional probability that a halo of virial mass 
$M_{h}$ contains $N$ QSOs above some specified (luminosity) threshold. A DM halo may contain zero, 
one, or more than one QSO. If more than one, the most massive galaxy is deemed to be at the centre 
of the potential well, and the less massive QSO(s) are the `satellites' in that halo (even though 
they may still be relatively massive, e.g. $\gtrsim 10^{11} M_{\odot}$ themselves). 

\citet{Richardson2012} present estimates of the 2PCF for QSOs, and interpret them with the 
HOD framework. In order to explain the small-scale clustering, the HOD model 
requires that a small fraction, $f_{\rm sat} = (7.4\pm1.4) \times 10^{-4}$, of the QSOs be satellites 
in DM haloes at $z\simeq1.4$. The median masses of the host haloes of central and satellite QSOs 
at these redshifts are constrained to be $M_{\rm cen} = 4.1^{+0.3}_{-0.4} \times 10^{12} \msolaroh$ 
and $M_{\rm sat} = 3.6^{+0.8}_{-1.0} \times 10^{14} \msolaroh$ respectively. Note that even thought centrals are expected to be the most massive QSOs in a given halo the satellites resides only in the massive haloes, whereas centrals QSOs can be found in relatively low mass haloes. Therefore, the median mass of host haloes of satellites QSOs is found to be two order of magnitude larger than central QSOs.

\citet{Shen2013} also present measurements of the 2PCF, this time via the cross-correlation of 
$\simeq\,$8,200 SDSS QSOs and $\simeq\,$350,000 massive red galaxies from the SDSS-III Baryonic Oscillation Spectroscopic Survey (BOSS) at $0.3 < z < 0.9$.  
They estimate a QSO linear bias of $b_{\rm Q} = 1.38\pm 0.10$ at $ \langle z \rangle = 0.53$ corresponding 
to a characteristic host halo mass of $~4 \times 10^{12} \msolaroh$ (compared with a characteristic host halo mass for galaxies of $~10^{13}\msolaroh$). Comparing these measurements 
with HOD models suggests that QSOs reside in a broad range of host haloes. The host halo mass 
distributions significantly overlap with each other for QSOs at different luminosities, implying 
a poor correlation between halo mass and instantaneous QSO luminosity. \citet{Shen2013} 
also find that the QSO HOD parameterisation is largely degenerate such that different HODs 
can reproduce the cross-correlations equally well, but with different satellite fractions 
and host halo mass distributions. 

\cite{2019MNRAS.487..275G} study the distribution of AGN host haloes using semi-empirical modelling. This model was shown to be consistent with current clustering measurements and found that AGNs host halo mass distribution is broad. They also predict that the fraction of satellite AGN increases towards the massive haloes.

The Sloan Digital Sky Survey (SDSS) has been the state-of-the-art in spectroscopic QSO surveys for the last 15 years. 
The Extended Baryon Oscillation Spectroscopic Survey (eBOSS) is the culmination of the SDSS-I, -II, -III and -IV 
quasar programmes and has recently completed a spectroscopic survey of $>$500,000 QSOs over 6,000 square degrees, 
covering redshifts $0.7 < z < 3.5$ \citep{lyke20a}. eBOSS is currently the premier dataset to measure QSO clustering \citep{ross20}. In near future eROSITA \citep{eROSITA} will provide the most promising X-ray AGN sample.

\citet{Eftekharzadeh2019} measure the quasar clustering signal across four orders of magnitude in scale, 
(\smash{$0.01 \lesssim r_{p} \lesssim 100 \mpcoh$}) at $z\simeq1.5$ using data from eBOSS. Using the HOD prescription, these authors find a satellite fraction of \smash{$f_{\rm sat} = 0.071^{+0.009}_{-0.004}$} and minimum mass of \smash{$M_{\rm min} = 2.31^{+0.41}_{-0.38} \times 10^{12} \msolaroh$} for the host dark matter haloes best describes the quasar clustering on all scales. \cite{2017MNRAS.468..728R} used a modified Sub-halo abundance matching method to model eBOSS QSOs showing the mean host halo mass of $5\times 10^{12}\msolaroh$. 


In this paper we extend these measurements of the clustering of galaxies and QSOs in eBOSS 
in order to:  
{\it (i)} understand the relation of the active QSOs to the general galaxy population; and 
{\it (ii)} understand the relation of the large-scale structure traced by the 
QSOs to the underlying DM halo distribution. 

We will make progress by employing and expanding on recent work using the `Multi-Tracer HOD' model
\citep[MTHOD:][hereafter Paper I]{2019arXiv191005095A} and apply this method to
the latest version (Data Release 16) of the SDSS-IV eBOSS data. Our
goal is to investigate a series of halo occupation distribution (HOD)
models that  described how QSOs populate the distribution of
dark matter haloes. We will use the luminous red galaxies (LRGs) and
the emission line galaxies (ELGs) that inhabit the same cosmological
volume as the QSOs to perform these tests and investigations. This
will allow us to discriminate between the models.

This paper is organized as follows. 
In Section 2, we present our data sample. 
In Section 3, we describe the techniques involved in our measuring the 2PCF, 
and note several effects that could give rise to systematics in the measurements.
In Section 4 we describe our Multi-tracer Halo Occupation Distribution (MTHOD) model 
and our Galaxy-QSO Occupation Distribution (GQOD) model that we use to model the galaxy and AGN population. 
In Section 5, we present our clustering measurements and the derived parameters.  
In Section 6 we place our new results in a broad context and note our main findings.
We conclude in Section 7.  
Appendix A gives technical details. 
We assumes a flat $\Lambda$CDM cosmology with $\Omega_m=0.307$, $\Omega_b=0.048$, $h=0.67$, $n_s=0.96$ and $\sigma_8=0.82$. Our assumed cosmology is close to the best fit parameters reported in  \citet{Planck2018} 
and motivated by the fiducial cosmology assumed in the $N$-body simulation \citet[MPDL2;][]{2012MNRAS.423.3018P} 
that we employ in our HOD models.

\section{Data}\label{sec:data}
In this section we describe the spectroscopic data from 
the SDSS eBOSS survey that we will use for our clustering 
measurements. We also will utilise new deep public imaging data 
from the Hyper Suprime-Cam (HSC) Subaru Strategic Programme. 

\begin{figure}
    \centering
    \includegraphics[width=0.45\textwidth]{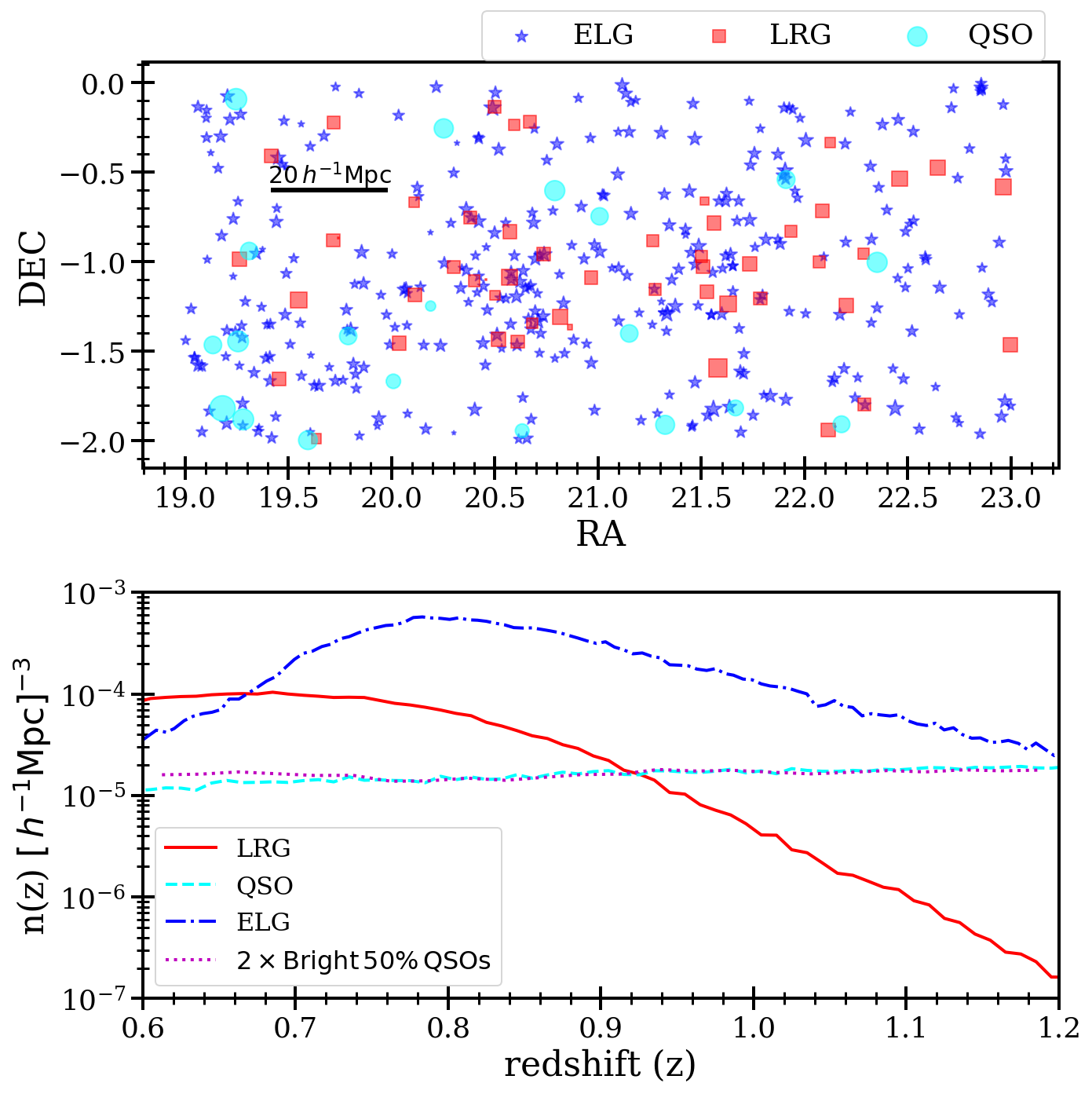}
    \caption{Sky coverage and number density distribution of our sample used in this paper. 
    The top panel shows a $\sim 4\times2 \deg^2$ patch of the eBOSS sample between redshifts
    0.82 and 0.88. The three tracers LRG, ELG and QSO are shown in red squares, blue stars 
    and cyan circles respectively. The varying symbol sizes represent the absolute AB $r$-magnitude 
    of each individual object. This clearly shows that QSOs are rare and bright compared to 
    LRG and ELG galaxies. The bottom panel shows the number density distribution of each tracer 
    in our sample. The red solid line, blue dotted-dashed line and cyan dashed line represents 
    LRG, ELG and QSO respectively. In this paper we apply a redshift cut at 0.7 and 1.1 for our analysis. 
    }
    \label{fig:coverage}
\end{figure}

\subsection{SDSS-IV: eBOSS}
We use data obtained from the extended Baryon Oscillation Spectroscopic Survey \citep[eBOSS:][]{Dawson2013} . This is one of the programmes of the wider 5 year Sloan Digital Sky Survey-IV \citep[SDSS-IV;][]{2017AJ....154...28B} using the BOSS spectrograph \citep{Smee2013} on the Sloan Telescope \citep{Gunn2006}.

The primary science goal of eBOSS was to measure the expansion of the Universe via LSS spectroscopic surveys. To achieve this, eBOSS comprises four different tracers: Luminous Red Galaxies (LRGs, with a redshift range $0.6<z<1.0$), Emission Line Galaxies (ELGs; $0.6 < z < 1.1$), Quasi-Stellar Objects (QSOs; $0.7 < z < 3.5$) and Lyman-$\alpha$ forest traced by QSOs ($z>2.1$). 

We use a subset of the eBOSS samples covering two fields between redshifts 0.7 and 1.1 where LRGs, ELGs and QSOs overlap in volume. The QSOs and LRGs sample the same area of the sky, but the ELG sky coverage is smaller. However, the ELG volume lies mostly inside the QSO/LRG volume by construction. Using the cross-correlation clustering technique, this overlap region can be also used to study the environemental dependence of QSOs, resulting in a sampling of the underlying dark matter distribution. We now briefly describe the relevant aspects of eBOSS sample selection, with more details being given by \citet{2016ApJS..224...34P} for LRGs, \citet{2017MNRAS.471.3955R} for ELGs and \citet{2015ApJS..221...27M} for QSOs. Table~\ref{tab:our_data} gives the number of objects from each selection used in this study. The total volume where all three tracers are observed is 0.64 $h^{-3}$ Gpc$^3$ between redshifts of 0.7 and 1.1.

\subsection{Luminous Red Galaxy (LRG) selection}
The LRGs are the most luminous and reddest galaxies, residing in massive dark matter haloes with high bias. The eBOSS LRGs are selected from SDSS imaging data \citep{2017ApJS..233...25A} in combination with infrared photometry from WISE \citep{Wright2010} using the following target selection rules:
\begin{align}
    r-i  & > 0.98 \label{LRG1} \\
    r-W1 & > 2(r-i) \label{LRG2} \\
    i-z  & > 0.625 \label{LRG3}
\end{align}
where $r$, $i$ and $z$ are the `model' magnitudes of the SDSS photometric bands and W1 refers to the WISE magnitude in the 3.4$\mu$m channel. The selections in equations \ref{LRG1}, \ref{LRG2} and \ref{LRG3} are designed to achieve the redshift range, reduce stellar contamination and reduce low-$z$ interlopers, respectively. The details of how these rules were derived and additional considerations are discussed in \citet{2016ApJS..224...34P}.

\subsection{Emission Line Galaxy (ELG) selection}
The eBOSS ELGs are expected to be star-forming galaxies at high redshift, and are thus selected based on high OII flux. The ELG sample is selected from DECAM Legacy survey \cite[DECaLS:][]{2019AJ....157..168D}. The target selection of ELGs in the North Galactic Cap (NGC) and South Galactic Cap (SGC), are slightly different due to availability of deeper data in SGC. We only use the SGC part of the ELG sample due to overlap with other tracers hence we only describe the SGC selection here.  The ELG selection has two parts; the first part is to select star-forming galaxies corresponding to OII emission lines using the $g$ band flux cuts
\begin{equation}
        21.825  < g < 22.825 .
\end{equation}
The second rule for ELG selection is to preferentially select galaxies around redshift 1.0 given by following equations: 
\begin{align}
    -0.068 (r-z) + 0.457 & < g-r < 0.112(r-z)+0.773\\
    0.218 (g-r) + 0.571 & < r-z < -0.555(g-r) + 1.901,
\end{align}
where $g,r,z$ are the observed magnitude in DECaLS $g$, $r$ and $z$ photometric bands. More details of how these rules were derived and additional considerations are discussed in \cite{2017MNRAS.471.3955R}.

\subsection{Quasar (QSO) selection}
\cite{2015ApJS..221...27M} describe in detail the requirements and how the eBOSS QSO sample is selected. First a supersample of QSOs is selected from SDSS imaging with either  $g<22$ or $r<22$ and $i_f>17$, where $g$ and $r$ are the PSF magnitude of SDSS photometric bands and $i_f$ is the FIBER2MAG. This supersample is passed through the XDQSOz algorithm \citep{2012ApJ...749...41B}, which assigns a probability for each object of being a QSO in a given redshift range. The eBOSS sample uses a probabilistic cut of $P_{\rm QSO}(z>0.9)>0.2$. An infrared cut in WISE imaging is also used to remove stellar contamination. The final QSOs sample with good redshifts is obtained using (IMATCH=1 or 2) along with a target completeness cut ($C_{\rm eBOSS}>0.5$) and spectroscopic completeness cut ($C_z>0.5$) as described in \cite{ross20}.

 \begin{table}
  \begin{centering}
    \begin{tabular}{l r r r}
      \hline
      \hline
   Sample & Number        & $n(z=0.86)$ & Overlap Area  \\
          & of objects    &              & with QSOs ($\mathrm{deg}^2$)\\
     \hline
     ELG  & 185,789 &   40 & 674\\
     LRG  & 98,086  &   10 & 4237 \\
     QSOs & 57,484  &    2 & 4808\\
      \hline
      \hline
    \end{tabular}
    \caption{The number of objects from each selection used in this study. 
    The approximate number density, in units of $10^{-5} (h^{-1}$ Mpc)$^{-3}$ is 
    also given. 
    }
    \label{tab:our_data}
  \end{centering}
\end{table}

\subsection{Quasar brightest 50\% sample}
We would also like to investigate how quasars populate their DM haloes as a function of QSO luminosity. QSO luminosity should depend on SMBH mass and mass accretion rate. With the link between SMBH mass and bulge mass established at low ($z\lesssim0.1$) redshift, and with a connection between bulge and halo mass, one might suspect that more luminous quasars (with more massive SMBHs) might populate their DM haloes in a different manner. 

We split the full QSO sample into the brightest 50\% of objects, as given by their $i-$band absolute PSF magnitude, noting that the observed $i$ band samples 3660-4530\AA\ rest-frame wavelength at $z=0.86$, where there are no strong broad emission lines. This is 28,742 objects which we call this sample the `Brightest 50\%'. 

The top panel of Figure~\ref{fig:coverage} shows a $\sim$~$4\times2 \deg^2$ patch of eBOSS between redshifts 0.82 and 0.88. The three tracers LRG, ELG and QSO are shown in red, blue and cyan coloured filled symbols, respectively. The size of the symbols represents the absolute $r$-band (AB) magnitude of individual objects. This clearly shows that QSOs are rare and bright compared to the LRGs and ELGs. The bottom panel of Figure~\ref{fig:coverage} shows the number density distribution of each tracer in our sample. In this paper we apply redshift cuts at $z=0.7$ and $z=1.1$ for our analysis. We note that around redshift $z=0.86$, the mean redshift of our measurements, the number density of LRG, ELG and QSO are $10^{-4}$, $4 \times 10^{-4}$ and $\smash{2 \times 10^{-5} \left[\mpcoh\right]^{-3}}$ respectively. The redshift distribution of the Brightest 50\% QSO sample is also shown in the bottom panel. The redshift distribution of the full QSO and Brightest 50\% QSO samples are the same, so a direct comparison between the two is reasonable.

\subsection{Hyper Suprime-Cam imaging}
Deep imaging data from Hyper Suprime-Cam \citep[HSC; ][]{Miyazaki2018} exist for a portion of our spectroscopic data. Imaging data from the the Hyper Suprime-Cam Subaru Strategic Programme (HSC-SPP)
cover part of the sky shown in Figure~\ref{fig:coverage}. 
Using the 8.2m Subaru Telescope \citep{Iye2004}, the HSC-SSP \citep{Aihara2018a, Aihara2019} 
currently offers the best combination of depth and image quality for a ground based survey. 
The Wide Layer achieves depths of {\it g}\,=\,26.6, {\it r}\,=\,26.2, {\it i}\,=\,26.2, {\it y}\,=\,25.3 and {\it z}\,=\,24.5
in the five broad-band filters. The seeing ranges from 0.58 to 0.77$''$ \citep{Aihara2019}. 
All data products are available for the Second Data Release at
\href{https://hsc-release.mtk.nao.ac.jp/doc/}{\url{https://hsc-release.mtk.nao.ac.jp/doc/}}.

The HSC-SSP data are of high enough quality to see galaxy groups out to $z\sim1$ \citep[e.g.][]{Umetsu2020}. 
Thus, we will use the HSC data to visually inspect the environments of the $z=0.7-1$ eBOSS QSOs. 
\section{Measurements and systematics}\label{sec:measurement}

\subsection{Clustering and the 2-Point correlation function}
Here we give a brief description of the 2PCF; for a more formal treatment the reader is referred to e.g. \citet{Peebles1980book}. 
The 2PCF, $\xi(r)$, is defined by the joint probability that two objects (e.g. galaxies) are found in the two volume elements $dV_{1}$ and $dV_{2}$ placed at separation $r$, 
\begin{equation}
 dP_{12} = n^{2} [ 1 + \xi(r) ] dV_{1} dV_{2}.     
\end{equation}
with $n$ being the object number density. 
To calculate $\xi(r)$, $N$ points are given inside a window $W$ of observation, which is a 3D body of volume $V(W)$. We calculate the position of each galaxy in 3-dimensional space by converting  the measured redshift to a line-of-sight distance using our fiducial cosmology. As usual, we also generate a catalogue of random points, with the same window function $W$ as the data, but without correlated positional information.

We then measure the galaxy auto-correlation function using the minimum variance Landay-Szalay estimator \citep{LandySzalay93} given by:
\begin{equation}
    \xi_{\rm auto}(\vec{r}) = \frac{DD(\vec{r})-2DR(\vec{r})+RR(\vec{r})}{RR(\vec{r})}.
    \label{eqn:landay-szalay} 
\end{equation}
where $DD$, $DR$ and $RR$ are the number of galaxy-galaxy, galaxy-random and random-random pairs as a function of vector separation in 3-dimensional space. The galaxy, QSO and random catalogues for the large-scale structure measurements for the SDSS DR16 sample are publicly available\footnote{\url{https://data.sdss.org/sas/dr16/eboss/}}.
The cross-correlations are measured using the following estimator: 
\begin{equation}
    \xi_{\rm cross}(\vec{r}) = \frac{D_1D_2(\vec{r})}{D_1R_2(\vec{r})}-1. \label{eqn:xicross}
\end{equation}
where $D_1D_2$ and $D_1R_2$ are the number of galaxy-galaxy and galaxy-random pairs from different samples. As is conventional, we project the 3-dimensional space onto a 2-dimensional space that decomposes pair separation vectors along the line-of-sight ($r_{\parallel}$) and perpendicular to the line-of-sight ($r_{\perp}$, or  $r_{p}$).
This gives us the 2-dimensional correlation function $\xi(r_{p}, r_\parallel)$. 
\begin{equation}
    w_{p}(r_{p}) =  \int^{r_2}_{r_1} \xi(r_{p}, r_{\parallel}) \, dr_{\parallel}.
\end{equation}
In practice, we measure the projected correlation ($w_p$) by integrating the 2-dimensional correlation function along the line-of-sight between $\smash{r_\parallel=-40\mpcoh}$ to $\smash{r_\parallel=+40\mpcoh}$ and using 25 equally spaced bins in logarithmic scale for $r_\perp$ between $0.1\mpcoh$ and $30\mpcoh$. Typically, the projected correlation function is integrated to $r_\parallel=100\mpcoh$ or larger to avoid the need to model redshift space clustering. But our model is evaluated in redshift space with full non-linearity and hence we do not have any constraint on minimum $r_\parallel$ for the projected correlation function. The projected correlation function $w_p(r_p)$ helps us constrain the HOD parameters that govern the galaxy-halo connection. To estimate the errors of our measurements, we create 86 jackknife regions for our sample and calculate the jackknife covariance of the $w_p$ measurements. Note that our jackknife region in the overlap sky area corresponds to approximately $6 \deg \times 1.3 \deg$. The $1.3\deg$ at the mean redshift of the sample corresponds to roughly $40\mpcoh$ and hence large enough for our measurements. 

\subsection{Potential systematic errors}
The clustering measurement is sensitive to the completeness of the observed galaxy sample for a given selection scheme. Therefore, it is important to account for variation in the number of detected galaxies as a function of their position in the sky, plus various selection biases due to the systematic errors introduced by instrumentation and measurement.

The number of detected galaxies and the spectroscopic success rate can be correlated with, for example, stellar density, extinction, sky brightness, airmass or position in the fibre plate \citep[see e.g., ][]{Bolton2012}. To remove these correlations, \cite{AshleyRoss12} introduced the use of systematic weights, and the use of systematic weights for LRGs, QSOs and ELGs is investigated by 
\cite{2018ApJ...863..110B}, \cite{2018MNRAS.477.1604G} and \cite{anand20}, respectively. 
We show in Paper I that measurement of \smash{$w_p$} at scales between $\simeq\,$1-30$\mpcoh$ is insensitive to the resulting corrections by any of the introduced systematic weights. We also note that due to fibre-collision the galaxy sample becomes highly incomplete below the fibre scale, which is approximately 65$''$ on the sky, corresponding to a scale of $\simeq 0.5 h^{-1}$ Mpc at $z=0.86$ \citep{2012MNRAS.427.3435A, 2012ApJ...756..127G, 2017MNRAS.472.1106B}. For the purpose of this paper we will not use the measurements or our models at scales smaller than the fibre collisions. 

\subsection{Environmental measures}
We also measure galaxy environment following the method described in \cite{2019MNRAS.483.4501A}. Here we briefly summarise the method. We focus on the measurement of local overdensity of galaxies around our sample.

In order to measure galaxy overdensity, we first create the Voronoi tesselation of the survey volume using the method developed in \cite{2019MNRAS.483.4501A}. This partitions the volume into disjoint cells each containing only one galaxy. We use the random catalogue provided with the Large Scale Structure catalogue and count the number of randoms in each Voronoi cell ($N^{\rm rand}_{\rm cell}$). We then estimate the density as the ratio of the mean density of randoms ($n_{\rm rand}$) to the random counts for each cell, and associate this with the galaxy in that cell, $\rho_{\rm cell}$: 
\begin{equation}
    \rho_{\rm cell} = \frac{n_{\rm rand}}{N^{\rm rand}_{\rm cell}} \label{eq:rhocell}. 
\end{equation}
The measured density is then smoothed at this chosen scale to determine the smoothed density ($\rho_{\rm smth}$) as 
\begin{equation}
\rho_{\rm smth} (\vec{r_0}) = \int \rho_{\rm cell}(\vec{r}) \mathcal{N}(\vec{r}-\vec{r_0},\sigma_{\rm smth}) \, d^3r \label{eq:rhosmth}\\
\end{equation}
with $\sigma_{\rm smth}=5\mpcoh$. This is then converted to an overdensity, $\delta_{\rm 5}$:
\begin{equation}
 \delta_{\rm 5} = \frac{\rho_{\rm smth}}{\bar{\rho}_{\rm smth}}-1 \label{eq:overdensity}.
\end{equation}
This allows us to assign a value of $\delta_5$ for each QSO and galaxy in the observed sample living in the overlapped volume. We finally measure QSO fraction as the function of $\delta_5$ using:
\begin{equation}
   f_{\rm QSO}(\delta)= \frac{C_{\rm QSO}(\delta;\Delta)}{C_{\rm LRG}(\delta;\Delta)+C_{\rm ELG}(\delta;\Delta)+C_{\rm QSO}(\delta;\Delta)} ,
    \label{eq:fQ}
\end{equation}
where $\delta=\log_{10}(1+\delta_5)$ and $C_{\rm tracer}(\delta;\Delta)$ gives the weighted count of number of object of a particular tracer with $\delta$ between $\delta-\Delta$ and $\delta+\Delta$. In these counts, we only consider objects with $-1<\delta<1$ and divide this in five bins with $\Delta=0.1$. We use only five bins in order to keep the size of the covariance matrix small while still having the overall trend of $f_{\rm QSO}$ with $\delta$. Each object in the sample is weighted such that the redshift distributions of all objects are the same.

\section{Modelling the galaxy and QSO Populations}
Our aim is to use the (cross-)clustering measurements 
of the LRG, ELG and QSO populations as constraints for 
our HOD models in order to understand the QSO population. 

We employ two HOD models: 
{\it (i)} the Multi-tracer Halo Occupation Distribution (MTHOD) 
model to model the overall galaxy population and {\it (ii)} 
the Galaxy-QSO Occupation Distribution (GQOD) model in order 
to model the statistical properties of QSOs as a distinct 
sub-population from the parent galaxy population.
The {\it anzatz} we use to model the QSO population is that the active galactic nuclei (AGN) observed as QSOs are {\it not special} in their inherent host galaxy properties, but are a  sub-sample of the global galaxy population. 


\subsection{MTHOD galaxy catalogues}
To model the galaxy population we use the Multi-tracer Halo Occupation Distribution (MTHOD) model and catalogue of \cite{2019arXiv191005095A}. The \mthod model introduces a new approach to model multiple tracers in the same volume. In general each of the tracers can have its own occupation recipe for the central and satellite galaxies. At the same time, the \mthod ensures that the joint occupation probabilities are well behaved by limiting the total probability of central galaxies in a halo to 1. It also forbids the non-physical situation of multiple types of galaxies at the centre of the same dark matter halo. The key parameters in \mthod models are the separate parameters for the occupation probability of central and satellite galaxies for each tracer; these are given in the Appendix.

The MTHOD mock galaxy catalogue is created using the MultiDark Planck simulation \citep[MPDL2: ][]{2012MNRAS.423.3018P} publicly available through the CosmoSim database\footnote{\url{https://www.cosmosim.org/cms/simulations/mdpl2/}}. MPDL2 is a dark matter only $N$-body simulation using the Gadget-2 algorithm \citep{2016MNRAS.457.4340K}. MDPL2 assumes a flat $\Lambda$CDM cosmology with $\Omega_m=0.307$, $\Omega_b=0.048$, $h=0.67$, $n_s=0.96$ and $\sigma_8=0.82$, and is a periodic box of side length 1000$\mpcoh$ sampled by $3840^3$ particles. A halo catalogue is generated using the ROCKSTAR\footnote{\url{https://bitbucket.org/gfcstanford/rockstar}} halo finder \citep{behroozi13} at an effective redshift of $z = 0.86$. 

The DM haloes are then populated using the following equations for central and satellite galaxies as a function of halo mass, $M_{\rm halo}$:
\begin{align}
     p_{\rm cen}^{\rm tot}(M_{\rm halo};\vec{\theta}) &=\sum_{\mathrm{tr}\, \epsilon \mathrm{TR}} p_{\rm cen}^{\rm tr}(M_{\rm halo};\theta^{\rm tr})\\
      \left< N_{\rm sat}^{\rm tot} \right>(M_{\rm halo}; \vec{\theta}) &=\sum_{\mathrm{tr} \,\epsilon \mathrm{TR}} \left< N_{\rm sat}^{\rm tr} \right>(M_{\rm halo};\theta^{\rm tr}),
\end{align}
where the sum is over all tracers in the list, ${\rm TR} = \left\{ \rm LRG,QSO,ELG \right\}$. This equation requires a constraint of $p_{\rm cen}^{\rm tot}\leq1$ for any halo mass. 
In paper I, all three tracers (i.e. LRGs, ELGs and QSOs) are modelled within the \mthod framework. However, in this paper, we take a different approach. We only use the LRG and ELG galaxies, and do not use the QSOs from the default model. Using the MTHOD mock catalogue, we can measure the clustering and the central/satellite properties for the LRG and ELG populations.

We assume that the \mthod galaxy catalogue models the complete set of galaxies that host eBOSS QSOs. The \mthod model samples galaxies starting from a minimum halo mass of $2.1 \times 10^{11} M_{\odot}$. The mean halo mass of eBOSS QSOs is shown to be $5 \times 10^{12} M_{\odot}$ \citep[see Figure 4 of ][]{2019arXiv191005095A}. The eBOSS galaxies thus cover the entire halo mass range needed to model the eBOSS QSO selection. 
Therefore, in the absence of any strong environmental effect or assembly bias, it is reasonable to assume that the \mthod galaxy catalogue models the complete set of galaxies that host eBOSS QSOs. We do expect this assumption to fail in detail \citep[as indicated by Figure 8 of ][]{2019arXiv191005095A} but this should be a second order effect given the current avaialble constraints on such effects

\begin{table*}
    \centering
    \caption{Parameters of Galaxy QSO Occupation Distribution (GQOD) models.}
    \begin{tabular}{p{1.2cm} p{10cm} p{2.5cm}} \hline\hline
    {\bf Parameters}             &  {\bf Description}                               & {\bf Used with } \\ [0.5ex] \hline\hline
      $f_{\rm on}(M_{\rm halo})$ & This models the probability of a galaxy to turn on with a given host halo mass. &                \\ 
                                 & This is modelled as a linear spline with 8 knots at locations                    & Inherent \& Enforced  \\
                                 & $\log_{10}(M_{\rm knots})=\left[11.2, 11.8, 12.15, 12.5, 12.75, 13.0, 13.5, 15.2\right]$ &   \\
                                 & It also require the constraint of $\int_{0}^{\infty} f_{\rm on}(M_{\rm halo}) n_{\rm gal}(M_{\rm halo}) dM_{\rm halo}= n_{\rm QSO}$  & \\
        & where $n_{\rm gal}(M_{\rm halo})$ and $n_{\rm QSO}$ are the number density of galaxies and QSO respectively. & \\
      [0.9ex] \hdashline \\
      & Probability of galaxy to turn into QSO given its host galaxy type & Inherent \& Enforced \\
     $G_{\rm type}$ &$\begin{array} {lcl} p(G_{\rm type}= {\rm LRG})  & = & f_{\rm LRG} \\ p(G_{\rm type}= {\rm ELG}) & = & 1-f_{\rm LRG} = f_{\rm ELG} \end{array}$ & \\ [0.9ex]\hdashline \\
     & Probability of galaxy to turn into QSO given its host galaxy position &   \\
     $G_{\rm pos}$ &$\begin{array} {lcl} p(G_{\rm pos}= {\rm satellite})  & = & f_{\rm sat} \\ p(G_{\rm pos}= {\rm central}) & = & 1-f_{\rm sat} \end{array}$ & Only for Enforced $f_{\rm sat}$ \\
     \hline\hline
    \end{tabular}
    \label{tab:GQOD_model_params}
\end{table*}
\subsection{Galaxy QSO Occupation Distribution (GQOD) model}
A second model, the Galaxy QSO Occupation Distribution (GQOD) model is employed to model the statistical properties of QSOs as a distinct sub-population from the parent MTHOD galaxy population. The probability of a galaxy being a QSO is given by: 
\begin{equation}
    P_{\rm QSO}(M_{\rm halo},G_{\rm type},G_{\rm pos})= f_{\rm on}(M_{\rm halo}) p(G_{\rm type})p(G_{\rm pos}),
    \label{eq:turn-on}
\end{equation}
where $f_{\rm on}$ is the probability for the galaxy to turn on with a given host halo mass, 
$p(G_{\rm type})$ is the probability for a galaxy to be a QSO given its host galaxy type and 
$p(G_{\rm pos})$ is the probability for a galaxy to be a QSO given its host galaxy position (central/satellite). 
A summary of these key parameters of the GQOD model are given in Table~\ref{tab:GQOD_model_params} and we do consider the three probabilities to be independent of each other. Here, ``turn on'' is shorthand for the physical processes involved having sufficient mass accretion in the QSO central engine that the QSO becomes luminous enough to be detected in our survey volume. 

We assume $f_{\rm on}$,  the fraction of galaxies that will have a QSO turn on, is a function of DM halo mass, $M_{\rm halo}$. This might have a wide distribution in halo mass \citep[see e.g., the deduced wide range of halo mass for central galaxies in][]{Richardson2012, ric13}. Alternatively, QSOs might reside in dark matter haloes of a certain particular mass range and hence the turn-on probability will have a narrow distribution with halo mass  \citep[see, e.g., ][]{KayoOguri2012,Eftekharzadeh2019}. We model $f_{\rm on}(M_{\rm halo})$ with a `linear spline sampling' of 8 halo masses between $10^{11}\msolaroh$ and $10^{15}\msolaroh$. 

Moving to $G_{\rm type}$ -- the probability that a host galaxy of a QSO may depend on host galaxy type --
we introduce the parameter $f_{\rm LRG}$ to define the fraction of QSOs with LRG host galaxies in the sample. We define $f_{\rm LRG}=1$ to mean that {\it only\/} LRGs can host QSOs; consequently, $f_{\rm LRG}=0$ will mean that only ELGs can be QSO hosts. If the posterior likelihood of either of these extremes is zero, then we can rule out the QSOs being turned on in only one type of galaxy. In a model where the QSO probability is independent of host galaxy type, the fraction of QSOs with a given type of host galaxy should be the same as the fraction of the host galaxy type in the parent population. 
$f_{\rm LRG}$ (and thus $f_{\rm ELG} = 1 - f_{\rm LRG})$ can be measured directly from data, as is done in \citet[][their Figure 7]{2019arXiv191005095A}. However, the match between the measured and modelled $f_{\rm LRG}$ and $f_{\rm ELG}$ is very close, so we are able to just use the smoother models. 

The probability a host galaxy is a QSO may also depend on `position', that is, whether it is a central or satellite galaxy. Studies including \citet{zhe09,Richardson2012,ric13} assumed different halo distributions for central and satellite galaxies, while \citet{KayoOguri2012} and \citet{Eftekharzadeh2019} assumed the halo to be indifferent in hosting a QSO as either a satellite or central galaxy. We include the parameter $G_{\rm pos}$ to encapsulate the positional information. 

We adopt two model flavours to study this aspect of QSO physics. In the first flavour, there is no dependence of QSO probability on whether the host galaxy is a central or satellite. Thus, the fraction of QSOs hosted by satellite galaxies will be equal to the fraction of satellite galaxies in the parent population. We call this an `inherent' satellite fraction and label it `$f_{\rm sat} \, (\rm inherent)$'. In the second flavour, the QSO probability can depend on whether the host galaxy is a satellite or a central and this is denoted by an additional parameter, $f_{\rm sat}$. This allows an extra degree of freedom where the satellite fraction in the QSO population does not have to represent the satellite fraction of the parent galaxy population. We call this an `enforced' satellite fraction, labelled `$f_{\rm sat} \, (\rm enforced)$'. 

The full list of parameters in our model is given in Table \ref{tab:GQOD_model_params}. We have a total of 9 parameters in the inherent $f_{\rm sat}$ model with 8 of them to model the halo mass dependence of $f_{\rm on}$ using linear spline with 8 knots listed in Table~\ref{tab:GQOD_model_params} and the 9th parameter is to model fraction of QSO with host galaxy as LRG ($f_{\rm LRG}$). The enforced $f_{\rm sat}$ model) has an additional 10th parameter to model the fraction of QSOs that are satellite galaxies $f_{\rm sat}$. 

Once we have simulated the QSO catalogue, we can predict the projected auto-correlation of quasars as well as the cross-correlation of quasars with LRGs and ELGs. Given the denser galaxy population one can also measure the fraction of QSOs as a function of overdensity. We use these four measurements to constrain the parameters of the model.

\subsection{Model constraints and parameter estimation}
For any point in the parameter space one can evaluate the probability of a galaxy to turn on using equation \ref{eq:turn-on} and hence obtain a sample of QSOs in the \mthod catalogue.
The LRG-auto, ELG-auto and LRG$\times$ELG cross clustering measurements are used to constrain the MTHOD model. Since these are presented in Paper I, we do not report them again here. The QSO-auto, QSO$\times$LRG and QSO$\times$ELG cross-correlation, and QSO fraction are the (new) measurements used to constrain the GQOD. 

We use {\tt emcee} \citep{2013PASP..125..306F} to sample the parameters for each of the two models in this work via a Markov chain Monte Carlo (MCMC) process. We then evaluate the auto and cross-correlation of QSOs with LRG and ELG samples using methods described in Section~\ref{sec:measurement}. This measurement requires pair counting to be performed at each step of sampling. We use the publicly available code {\tt corrfunc} \citep{manodeep_sinha_2016_55161} to evaluate pair-counts efficiently at each iteration. We also pre-compute the overdensity field for each galaxy in the \mthod catalogue using the method described in \cite{2019MNRAS.483.4501A} -- which is then used to evaluate the fraction of QSOs as a function of envoronmental overdensity. We developed a python library to efficiently create QSO samples as a fuction of our model parameters\footnote{\url{https://www.roe.ac.uk/~salam/GQOD/}}. In this process we use the number density of the QSO sample, treated as fixed at the mean redshift.

We use $w_p^{q}(r_p)$, $w_p^{qL}(r_p)$, $w_p^{qE}(r_p)$ and $f_{\rm QSO}(\delta)$ to denote the QSO auto-correlation, QSO$\times$LRG and QSO$\times$ELG cross-correlation, and QSO fraction, respectively. The data and model are denoted by $\vec{D},\vec{M}(\theta)$ respectively.
\begin{align}
    \vec{D}         &=[w_p^{q}(r_p),        \, w_p^{qL}(r_p),         \, w_p^{qE}(r_p),        \, f_{\rm QSO}(\delta)] \\
    \vec{M}(\theta) &=[w_p^{q}(r_p,\theta), \, w_p^{qL}(r_p,\theta),  \, w_p^{qE}(r_p,\theta), \, f_{\rm QSO}(\delta)].
    \label{eq:Data_and_Model}
\end{align}
We then evaluate a model $\chi^2$ using following equation:
\begin{equation}
    \chi^2 (\theta) = (\vec{D}-\vec{M}(\theta)) \, C^{-1}(\vec{D}-\vec{M}(\theta))^T ,\\    
    \label{eq:chi2}
\end{equation}
where $\theta$ represents sampling parameters and $C^{-1}$ is the inverse covariance matrix obtained from jackknife analysis including the errors in estimate of model prediction. The MCMC process then samples the parameters according to the $\chi^2$ given by the model.

\begin{figure*}
    \centering
    \includegraphics[width=1.0\textwidth]{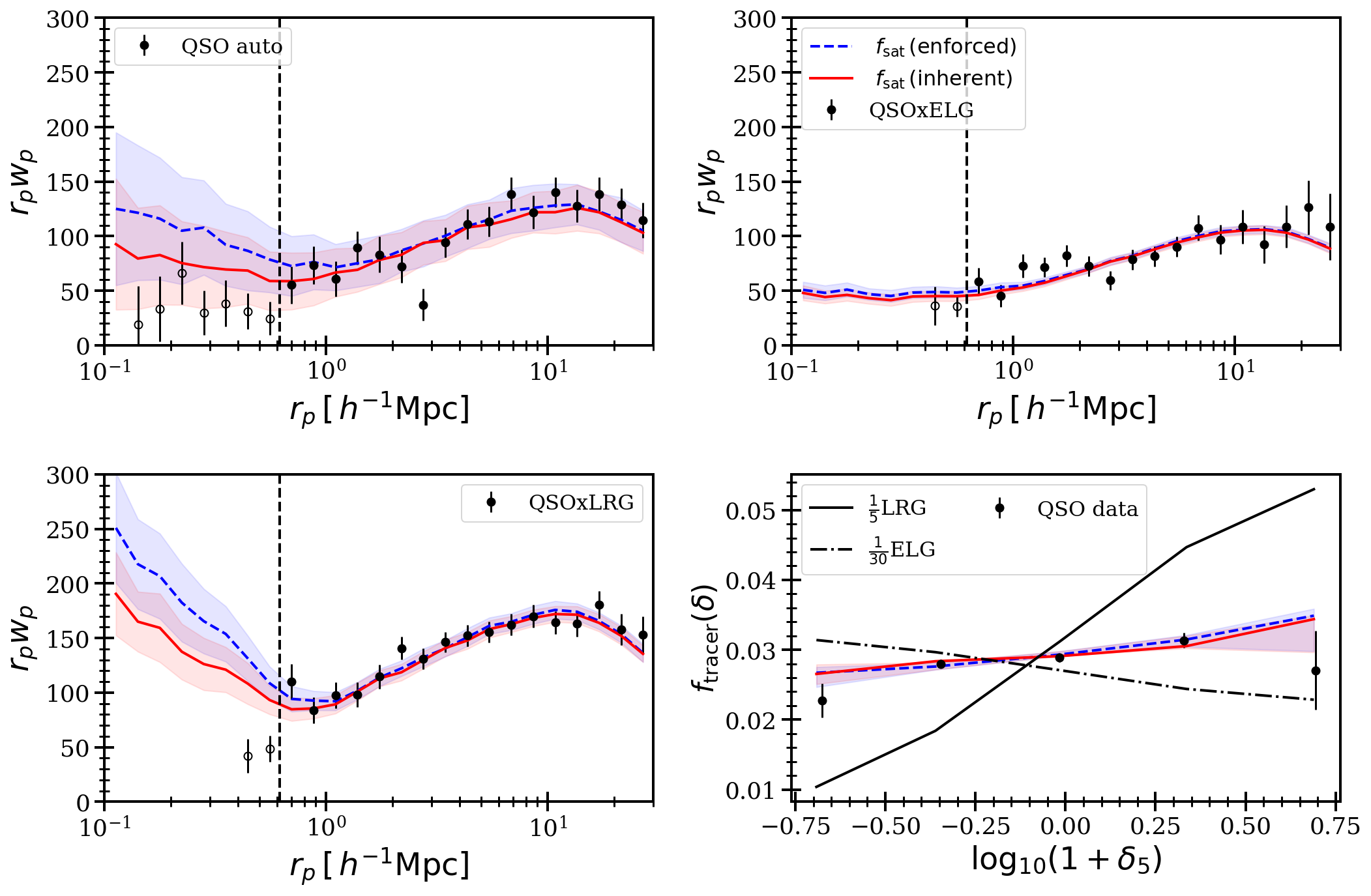}
    \caption{{\it Top Left}, {\it Top Right} and {\it Bottom Left} panels shows the projected correlation function ($w_p$) for the QSO auto- , the QSO$\times$ELG  and QSO$\times$LRG cross-correlation functions, respectively. The black points are the measurements from eBOSS along with the jackknife errors. The dashed vertical black line represents the fibre collision scale and data points below this are given as open circles (and not used in the analysis). {\it Bottom Right:}  shows the QSO fraction as a function of environmental overdensity. For comparison, we also show the fraction of LRGs ($f_{\rm LRG}$) and ELG ($f_{\rm ELG} = 1 - f_{\rm LRG}$) with overdensity by black solid and black dashed-dotted lines, scale by the factors given in the legend. In all panels,  the blue dashed line shows the model fit with an enforced satellite fraction as additional free parameter (the shaded region giving 1$\sigma$ errors). The red line shows the best-fit model where satellite and central galaxies are equally likely to host a QSO.}
    \label{fig:wp-auto-cross}
\end{figure*}

\section{Results}\label{sec:result}
In this section we present our results. Our convention is to report data 
measurements with black data points. The two flavours of the GQOD model will be 
represented by red (solid) and blue (dashed) lines for the `inherent' and 
`enforced' model fits, respectively.

\subsection{Clustering results}
We analyse the eBOSS QSO auto-correlation and the QSO-LRG and QSO-ELG cross-correlations between redshifts of $z=0.7-1.1$. The measurement of these statistics along with the best-fit GQOD models is shown in Figure~\ref{fig:wp-auto-cross}. 
The top-left, top-right and bottom-left panels of Figure~\ref{fig:wp-auto-cross} show the QSO auto-correlation,
\smash{$w_p^{q}(r_p)$},  the QSO$\times$ELG cross-correlation \smash{$w_p^{qE}(r_p)$}, and the QSO$\times$LRG cross-correlation \smash{$w_p^{qL}(r_p)$} functions, respectively. 
The measurement from eBOSS data is shown with black circle with error bars estimated from the jackknife sampling method. 
The vertical black dashed line shows the fibre collision scale at the mean redshift ($z=0.86$) and the projected correlation function below this scale are shown with open circle and not used in the analysis. 
The dashed blue and solid red lines with shaded region shows the enforced and inherent $f_{\rm sat}$ models respectively with its $1\sigma$ constraint in all panels. 
Note that different scales are correlated which is accounted for in our full covariance matrix.

The models describe the data very well in the range of scale considered in the current analysis. The smaller 1$\sigma$ regions in the \smash{$w_p^{qE}$} QSO$\times$ELG correlation function is due to the higher $n_{\rm gal}$ number density of the ELG sample. Since we do not sample the smaller (`1-halo' term) scales, there is no discernible difference in the models whether or not the satellite fraction is an additional free parameter.

\subsection{Environment and QSO fractions}
We  measure the QSO fraction as a function of galaxy overdensity environment. 
Recall, $f_{\rm QSO}(\delta)$ is measured straight from the data using the Voronoi cell estimation method (Section 3.3). This is presented in the bottom-right panel of Figure~\ref{fig:wp-auto-cross}.

The QSO fraction is around 3\% across the whole range of environments sampled (black points in Figure~\ref{fig:wp-auto-cross}, bottom right). The best-fit models (red and blue lines with 1$\sigma$ shadings)  again fit the data very well. The models rise monotonically from around 2.6\% for the least-dense regions to around 3.4\% for the most dense environments. 

In Figure~\ref{fig:wp-auto-cross}, we also show the LRG fraction, $f_{\rm LRG}$, as the solid (black) line and the ELG fraction (where $f_{\rm ELG} = 1 - f_{\rm LRG}$), as the dashed-dotted (black) line for comparison. The behaviour of the LRG and ELG fraction with overdensity is probably dominated by galaxy quenching being more efficient in the high halo mass regime \citep{2019arXiv191005095A}, hence leading to the high LRG fraction in the most overdense regions. We note that the QSO fraction has essentially a flat dependence on overdensity. This is in contrast to both the LRG or ELG fractions, which have an environmental density dependence. This implies that the QSO population must contain a mixture of LRGs and ELGs as host galaxies.

\begin{figure}
    \centering
    \includegraphics[width=0.5\textwidth]{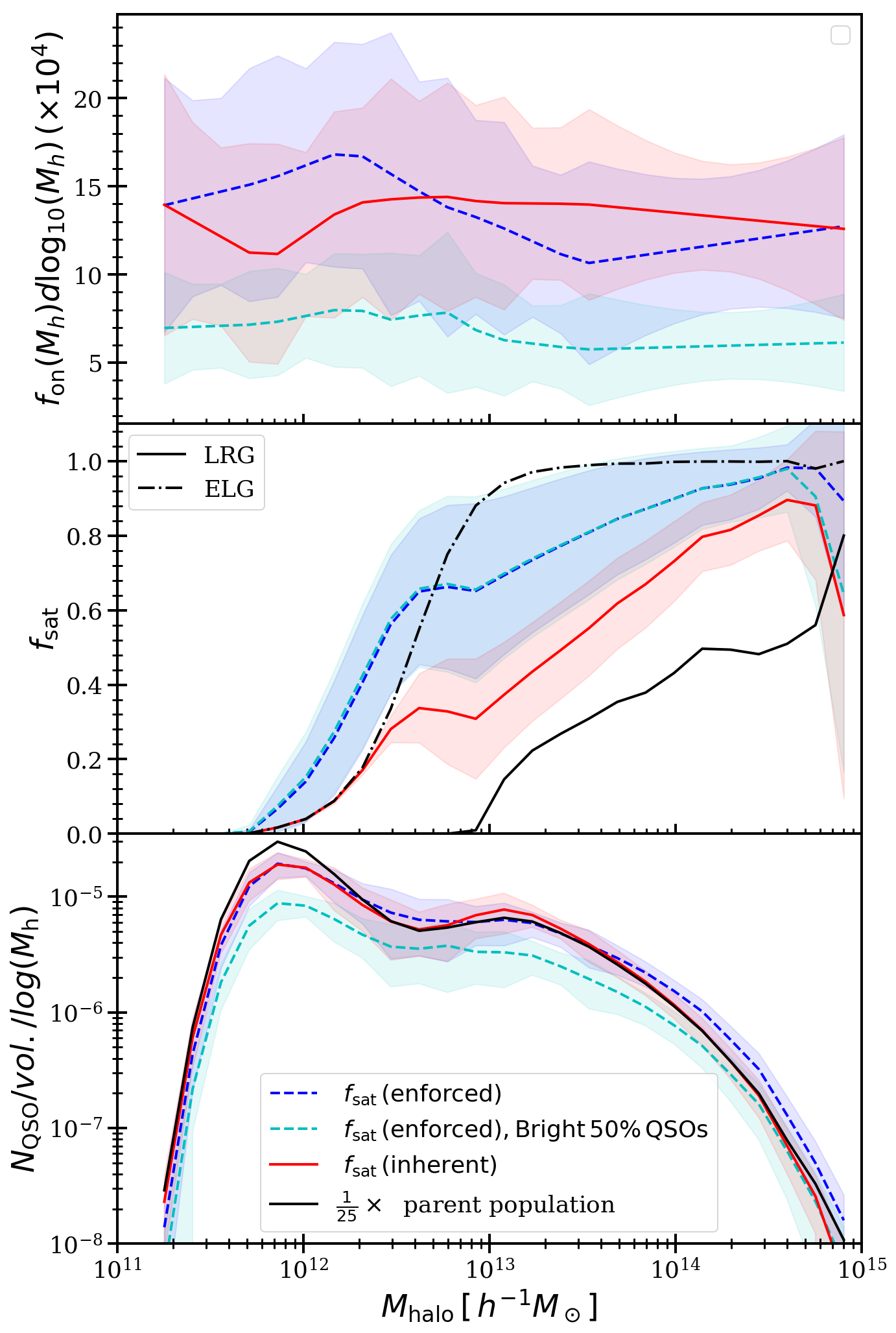}
    \caption{The top, middle and bottom panels show $f_{\rm on}$, $f_{\rm sat}$ and number density of QSOs per unit logarithm of halo mass as the function of halo mass respectively. In all panels the dashed blue and solid red lines show the two models with enforced $f_{\rm sat}$ and inherent $f_{\rm sat}$ respectively with shaded regions representing 1$\sigma$ errors. The $f_{\rm on}$ in the top panel presents the probability of a galaxy to turn on, which is also called the duty cycle of the QSO. The middle panel shows the fraction of QSOs living in satellite galaxies. For comparison, we also show in this panel the fraction of satellites for LRGs and ELGs, using black solid and dashed-dotted line respectively. The bottom panel shows number density of QSOs per unit $\log_{10}(M_h)$ and the black solid line shows the distribution for the parent galaxy population, scaled down by a factor of 25 for comparison.}
    \label{fig:Halo_occupation}
\end{figure}

\subsection{Dependence on DM halo mass}
Figure~\ref{fig:Halo_occupation} shows the best-fit GQOD values derived for 
$f_{\rm on}$ (the fraction of galaxies that will have a QSO turned on), 
the fraction of QSOs that are satellites ($f_{\rm sat}$) and the 
number density of QSOs, $N_{\rm QSO}$, as a function of DM halo mass. 
As before, the best-fit $f_{\rm sat}$ (inherent) model is given by 
the solid (red) line, with the  $f_{\rm sat}$ (Enforced) model given 
by the dashed (blue) line. $1\sigma$ errors are given by the shaded regions. 

In the top panel of Figure~\ref{fig:Halo_occupation} we see the fraction of galaxies that have a QSO turned on is essentially independent of halo mass for both the Enforced or inherent models, i.e., 
the probability that a galaxy has quasar activity is independent of halo mass. 
This is a key result: it implies that the halo mass distribution of QSOs is very broad, despite the model having freedom to choose a narrow range of halo mass through our spline fit. In order to test that the flat nature of $f_{\rm on}$ is indeed a better model we tested a lognormal model for $f_{\rm on}$. We looked at $\chi^2$ value in a grid of mean halo mass and scatter for lognormal model finding the minimum $\chi^2/{\rm dof}=72/53$ for a model with mean halo mas of $10^{13} \msolaroh$ with width 1.05. We also observed that the smaller width is strongly ruled out by having large value of $\chi^2$ whereas larger width increases $\chi^2$ marginally. When this is compared with our default model giving best fit with $\chi^2/{\rm dof}=42/45$ then we can be more confident that the flat $f_{\rm on}$ describes the data very well.

The middle panel shows the satellite fraction as a function of halo mass. 
The red line shows the satellite fraction for the best-fit inherent $f_{\rm sat}$ model, while the blue line shows the satellite fraction for the best-fit enforced $f_{\rm sat}$ model. 
The fraction of satellites being QSOs rises from 0\% at masses above $\simeq\,$5$\times10^{12} M_{\odot}$.
Around $\simeq\,$2$\times10^{13} M_{\odot}$. Both models put the fraction of QSOs being in satellites galaxies at over 50\%. 
For the dashed blue line that represents the satellite fraction in the enforced $f_{\rm sat}$, the model slightly prefers the QSOs to turn on in satellite galaxies (more notable at lower halo mass) leading the blue dashed line to always be above the red dashed line
(see section ~\ref{sec:discussion_fsat} for more discussion). \cite{2019MNRAS.487..275G} estimate the satellite fraction for AGN as the function of X-ray luminosity finding it to be around 10-20\% consistent with our satellite fraction for eBOSS QSOs.

In this middle panel, we also show for comparison the fraction of satellites for LRGs and ELGs with solid black and dashed dotted lines. There are no satellite galaxies that are LRGs in haloes of  
$M_{\rm halo} < 10^{13} M_{\odot}$ \citep[as found in previous LRG HOD studies: e.g.][]{Zehavi2005b, Ross2008, ReidSpergel2009, White2011}. In haloes of $10^{13} M_{\odot} < M_{\rm halo} < 10^{14.5} M_{\odot}$, the fraction of satellites that are LRGs increases from 0 to just under 60\%. Then, for haloes with $M_{\rm halo} \gtsim 10^{14.5} M_{\odot}$ the fraction of satellites that are LRGs jumps to near 80\%. This is because a halo can only have one central but more massive haloes can have multiple objects defined as satellites (e.g. in groups and clusters); thus the satellite fraction can, and will, approach 100\%. The reason $f_{\rm sat}$ levels at $\simeq\,$60\% for the LRGs is because we have many more satellites that are ELGs; in a model without ELGs, then by default the satellite fraction for LRGs would be 100\% in the most massive haloes. 



The bottom panel shows the number density of QSOs per unit logarithmic halo mass. The black solid line shows the same distribution for the parent population scaled down by a factor of 25. We note that the halo mass distribution of QSOs is very similar to that of the host population, which comes from the fact that the turn-on probability of QSOs is independent of halo mass. We note that this predicts a very broad distribution of QSO host DM halo mass. Overall, quasars inhabit dark matter haloes in essentially an identical way to the full galaxy population, although they are not as common.

\subsection{QSO dependence on luminosity}

In Figure~\ref{fig:Halo_occupation} we also show, with the cyan line and shading, 
the best-fit GQOD values derived for the  enforced $f_{\rm sat}$ model, but 
only using the brightest 50\% of the QSO data. 

Using the Brightest 50\% data, and refitting the best-fit models, we see from Figure~\ref{fig:Halo_occupation} in the top panel that the probability for a galaxy to have quasar activity is $\simeq\,$50\% for the brightest half of the quasar population, but remains independent of halo mass. The amplitude of the model fit simply comes from the number density; indeed, if one were to reduce the QSO population via a random sampling fraction $x$, then the probability of a galaxy having quasar activity will also reduce by a factor $x$. The key thing to note is this function remains flat, i.e. independent of halo mass. 

From Figure~\ref{fig:Halo_occupation} and the cyan line in the middle panel (which is very similar to, but under the blue line/shade), the fraction of satellite galaxies that are quasars is independent of quasar luminosity. 

Finally, in the bottom panel of Figure~\ref{fig:Halo_occupation}, we see that 
 overall, the 50\% more luminous quasars inhabit dark matter haloes, as function of mass, in essentially an identical way to   the full QSO population. As stated above, the probability of turn on for the smaller number of more luminous QSOs is reduced,     though we have same number of galaxies in our parent sample.

\begin{figure}
    \centering
    \includegraphics[width=0.5\textwidth]{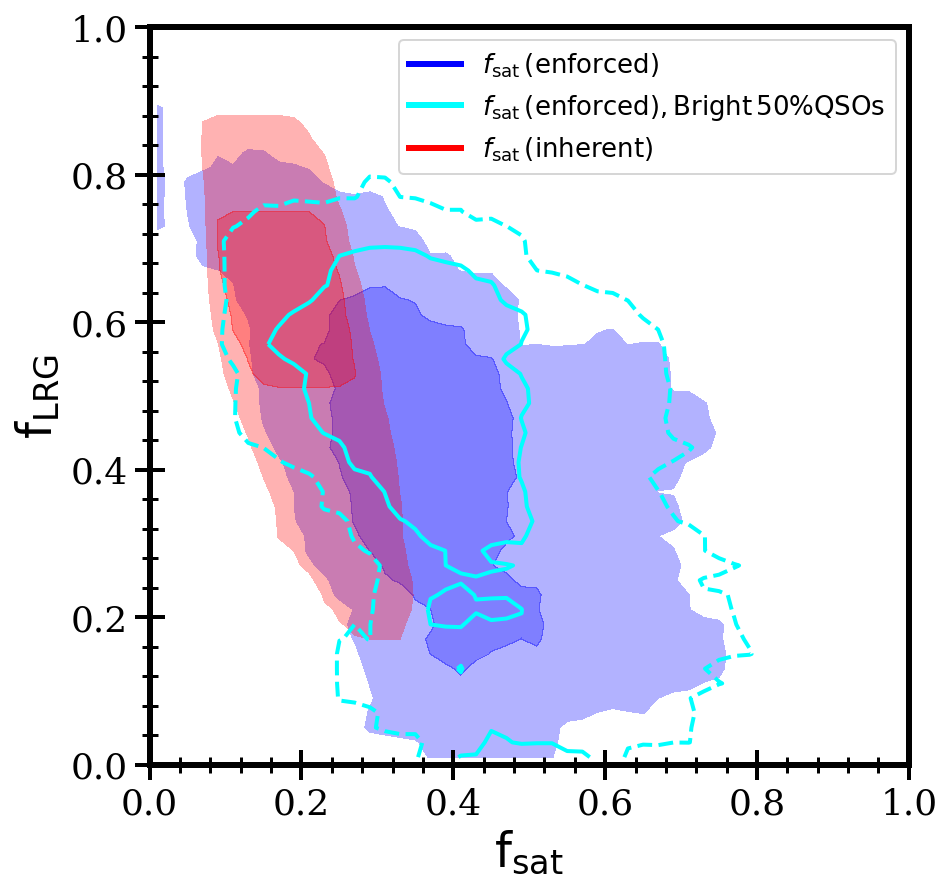}
    \caption{This shows the two-dimensional posterior of the fraction of satellites in QSOs ($f_{\rm sat}$) and the fraction of LRGs in QSOs ($f_{\rm LRG}$) for the two models. We note that for the red contours $f_{sat}$ is a derived parameter, whereas forthe  blue contours it is a free parameter. This shows that QSOs cannot come entirely from either LRGs or ELGs but rather are a mixture with roughly equal proportions. We also note that QSOs could have significantly larger satellite fractions and in the blue contour where we allow satellites to have non-equal probabilities of converting to QSOs then we find that the data prefer even larger satellite fractions. }
    \label{fig:fsat-fLRG-2d}
\end{figure}

\begin{figure}
    \centering
    \includegraphics[width=0.48\textwidth]{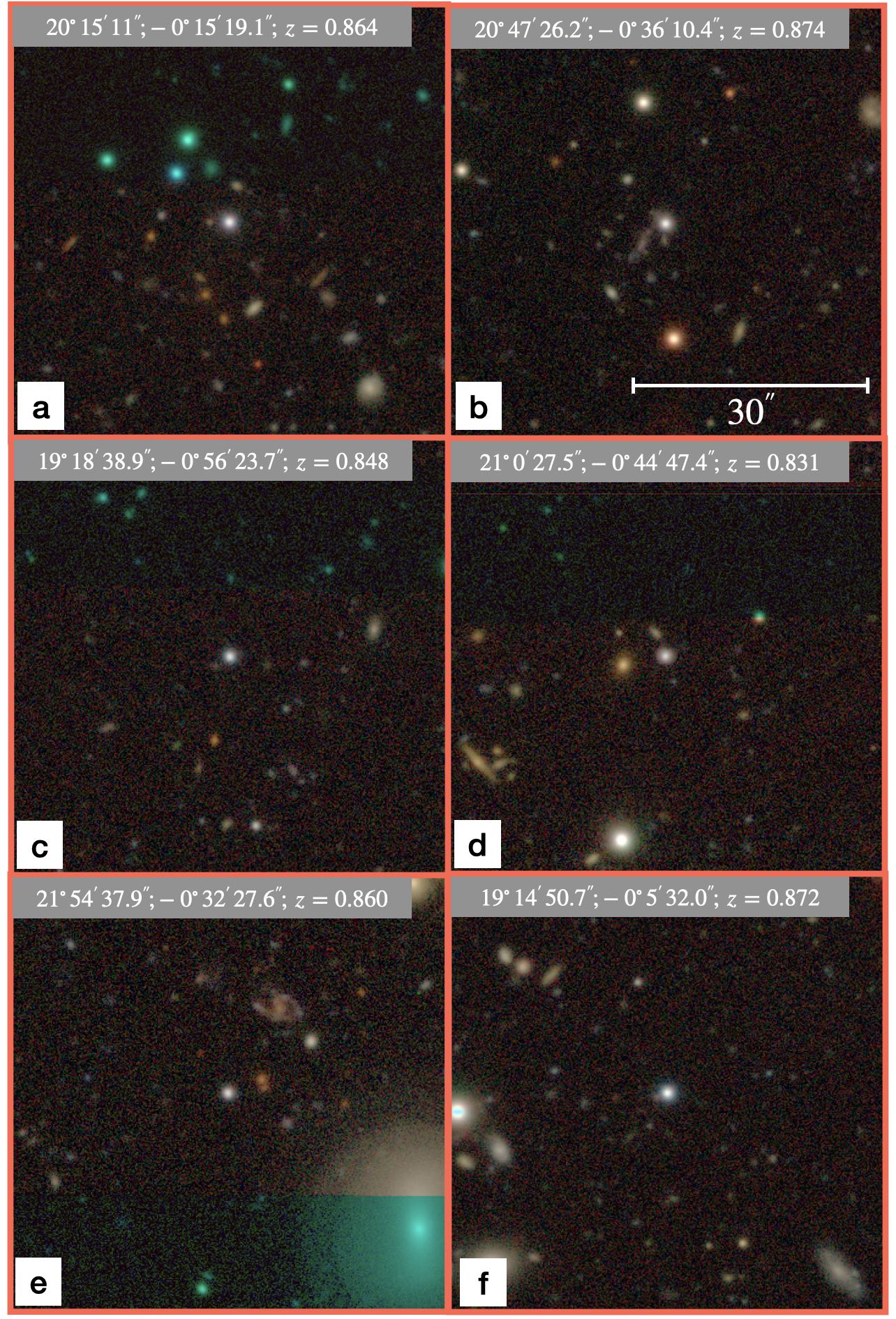}
    \caption{Cutout images from HSC data release 2 around six different QSO randomly selected in our sample. Each of the cutouts is centred around an eBOSS QSO, with a field of view about $1'$ across. This shows a sampling of the different environments that QSOs inhabit. The cutouts were created using the HSC 
    public data access tools.}
    \label{fig:HSC-cutout}
\end{figure}

\subsection{QSO dependence on galaxy type and position}
Figure \ref{fig:fsat-fLRG-2d} shows the two dimensional posterior distribution of the fraction of LRGs ($f_{\rm LRG}$) and the fraction of satellites ($f_{\rm sat}$) in the QSO population. 
The red and blue contours are the two model satellite fractions and the dark and light colour regions show $1\sigma$ and $2\sigma$ constraints. 

Consider the inherent model (red contours): since the models do not contain $f_{\rm LRG}=1$ or $f_{\rm LRG}=0$, this model rules out (at the $>3\sigma$ level) the possibility that the QSO population comes entirely from either the LRG or ELG population (when we do not allow the extra degree of freedom in the inherent satellite fraction model). 
It also shows that roughly $20\%$ of QSO host galaxies are satellite galaxies with about $60-70\%$ of QSO hosts being an LRG galaxy. 

When we allow the additional freedom from satellites in the enforced $f_{\rm sat}$ model (shown in blue), we are still able to rule out the possibility of the LRGs hosting the eBOSS QSO population at the $3\sigma$ level; but QSO host galaxies being entirely ELGs is possible (at the 2$\sigma$ level). Allowing the additional freedom in the model, the enforced $f_{\rm sat}$ prefers a larger fraction, $\sim$40\%, of QSO hosts to be satellite galaxies, but a lower fraction ($20-60\%$) of LRG host galaxy (compared to the inherent model).  This is due to both the LRGs and satellites residing in haloes with relatively higher halo mass. Therefore one can get the same clustering by either having a large LRG fraction with lower satellite fraction as in the inherent model, as by having a large satellite fraction with a smaller LRG fraction (as is the case for the enforced model). 

It is also interesting to note that in either case the satellite fraction between $20\%-40\%$ is much larger than host galaxy population and disagrees with other analysis of QSO satellite fraction \citep{2011ApJ...741...15S,Richardson2012}. Generally it is considered that large satellite fraction will mean enhanced clustering within the 1-halo term and hence our large satellite fraction might mean we overpredict the QSO auto-correlation at scales below $1\mpcoh$. But this is not the case, as shown in the top left panel of Figure \ref{fig:wp-auto-cross} where our model is consistent with the observed QSO clustering at these smaller scales. This is due to the QSO number density being very low; thus, even when we allow the model and QSOs to have a large satellite fraction, there will only be a few QSOs per halo and hence the central-satellite or satellite-satellite terms will not be as strong. 

\subsubsection{HSC Imaging of QSO Groups}
Our results suggest that a large percentage of QSO host galaxies are satellites. 
We turn to the deep HSC imaging data to see if we can find direct examples of this. 

A collage of HSC cutouts centred on six QSOs from the eBOSS sample is shown in Figure~\ref{fig:HSC-cutout}. 
The $RGI$ filters were mapped to a RGB colour scheme using the HSC public data access tools\footnote{\url{https://hsc-gitlab.mtk.nao.ac.jp/ssp-software/data-access-tools}}. 

The six QSOs are the ones shown in the top panel of Figure~\ref{fig:coverage} with a declination cut $\delta>-1.0^{\circ}$ to restrict to the region with HSC imaging. 
The field of view is $\simeq\,30''$ on a side, corresponding to a scale length of $\simeq\,$240\,kpc 
at $z=0.86$ for our given cosmology. 

In the collage we see several different example  environments for the QSO. 
Note, the QSO appears in the centre of the image; this does {\it not\/} indicate at all that the
QSO lies the centre of the group, or at the peak of the DM halo. 
From Figure~\ref{fig:HSC-cutout}, we see: 
in panel {\it (a)}, the top left panel, the QSO in the outskirts of a Red galaxy group; 
in panel {\it (b)}, top right, the QSO in the outskirts of a Blue galaxy group;
in panel {\it (c)}, middle left, the QSO in the centre, basically on its own; 
in panel {\it (d)}, middle right have the QSO in a pair with an LRG; 
in panel {\it (e)}, bottom left, the QSO is possibly in the outskirts of a galaxy group again, potentially with a massive `Red Spiral' as the central galaxy (to the top right, `1 o'clock position' of the QSO); 
and finally in panel {\it (f)}, bottom right, we show a QSO on its own again, but 
possibly the brightest object in a small group. 

Without having spectroscopic redshifts for the objects in these cutouts, it is of course very tricky to 
confirm galaxy group membership. However, given the depth and seeing quality of these images, and the lack of obvious foreground galaxies, it is entirely reasonable to assume we are seeing the environments of the QSOs.


\section{Discussion}
In this section, we discuss our key results and their implications.
We first discuss how the QSO population traces the underlying 
dark matter halo mass function (Section~\ref{sec:discussion_DMH}). 
We then place our results for the QSO satellite fraction in context, 
comparing with previous studies in Section~\ref{sec:discussion_fsat}.
In Section~\ref{sec:discussion_lrgs} we think about the inferences that may be made from
our clustering results regarding the growth history of stellar and black hole mass.
In Section~\ref{sec:discussion_implications} we note the immediate implications of 
this current work and point towards future investigations.

\subsection{QSOs and the dark matter halo mass function}\label{sec:discussion_DMH}

One key motivation for this study was to answer the question: how do QSOs populate their 
dark matter haloes as a function of mass? Figure~\ref{fig:Halo_occupation} shows us the 
answer here: the parameter $f_{\rm on}$ is flat, and so the fraction of galaxies 
that have a QSO turned on is essentially independent of halo mass for both the enforced or 
inherent models. {\it The probability that a galaxy has quasar activity is therefore independent of halo mass.} 
The result in the bottom panel of Figure~\ref{fig:Halo_occupation} immediately follows: 
the QSOs sample the same DM halo mass function as the parent galaxy sample. 
\citet{ConroyWhite2013} find the same result. In fact, their model is consistent with 
QSOs being equally likely to exist in galaxies, and therefore dark matter haloes, over a wide 
range in masses and suggests a single QSO duty cycle at redshift $z<3$. 

In fact, this observation has been already recognised in the X-ray community
\citep[e.g.][]{Leauthaud2015, Mendez2016, Powell2018, Plionis2018,  Krishnan2020}.
All these studies find no significant differences in the 
clustering properties of X-ray AGN compared to a matched galaxy sample. That is, X-ray AGN inhabit DM haloes that are consistent on average with the overall 
inactive galaxy population. This result runs from the local Universe \citep[$0.01<z<0.1$; ][]{Powell2018} 
to high redshifts, $z\simeq4.5$ \citep[][]{Krishnan2020}. 
Studying narrow-line AGN, and comparing to a matched control samples of inactive galaxies,  \citet{Li2006c} find that AGN have almost the same clustering amplitude as the control galaxies, on scales larger than a few Mpc.  
Here we are showing, {\it for  the first time}, that the same is true for the blue optically selected broadline QSOs at modest redshift.  

For the Brighter 50\% QSO sample, we find that $f_{\rm on}$ remains flat, and thus this 
sample also directly maps to the same DM halo mass function as the parent galaxy sample. 
This is another key result as it immediately explains the lack of dependence on luminosity 
for QSO clustering \citep[e.g.][]{daAngela2008, Shen2009, Shen2013, Chehade2016, Powell2020}.  This is consistent with the result that the QSO luminosity possibly has large scatter at fixed black hole mass and hence do not particularly correlates with the host halo mass \cite{2014ApJ...782....9H}.

\subsection{The fraction of QSOs that are satellite galaxies}\label{sec:discussion_fsat}
We remind the reader that our estimate of the fraction of QSOs that are in
satellite galaxies comes from our GQOD model and the MTHOD mock catalogue:
we derive the QSO satellite fraction, $f_{\rm sat}$, by comparing the model to
the data. Our results suggest that above a halo mass of $\simeq\,10^{12} M_{\odot}$ 
a substantial number of QSOs can be found in satellite galaxies. 

A range of previous studies give a very large range of measured satellite fractions for QSOs. 
\citet{Richardson2012} report a satellite fraction for $z\simeq1.4$ QSOs as 
$f_{\rm sat} = (7.4\pm1.4)\times10^{-4}$. This is a much smaller figure than our large percentage, though it is not straightforward to compare since \citet{Richardson2012} quote a probability density function of the satellite fraction as given by all their HOD models, whereas we have the satellite fraction as a function of DM halo mass. Also, and using very similar data to \citet{Richardson2012}, \citet{KayoOguri2012} find  $f_{\rm sat} = 0.054^{+0.017}_{-0.016}$, i.e. nearly an order of magnitude smaller than our figure of >30\% at $M_{\rm halo} \gtrsim3\times10^{12} M_{\odot}$.
This apparent discrepancy in satellite fraction can possibly be explained by realizing that these studies use small-scale pairs built from samples of binary quasars. If, for instance, binary quasars consisted of two QSOs that had very similar masses, then it is not clear that one being a satellite and the other being a central is a meaningful distinction. In this scenario, on average, a "non-central" member of each binary quasar could be interpreted as part of a small fraction of high-mass satellites. 
 We do note that we do not sample the 1-halo term very well (due to fiber collisions), and so suggest that our sample samples quasars in e.g. groups where the QSO can be a satellite, and this occupation fraction might be (very) different from e.g. binary QSOs in the same halo but potentially different sub-haloes. 

\cite{Starikova2011} present results showing that the Chandra/Bo\"otes AGN are predominantly 
located at the centres of dark matter haloes and tend to avoid satellite galaxies in haloes of this or higher mass. However, also using moderate luminosity X-ray AGN (at $z < 1$ from the COSMOS field), \citet{Leauthaud2015} report a mean satellite fraction of 
$\left \langle f_{\rm sat} \right \rangle = 18\pm2$ \%. 

\citet{WangLi2019} is another key result here. Using a sample of 100,000 AGN from SDSS \citep[with the AGN being classified via the BPT diagram, ][and lying mostly below $z = 0.3$]{BPT}, these authors also perform a clustering measurement in order to investigate the DM halo properties of narrow-line AGN. \citet{WangLi2019} investigate the central/satellite fraction for the AGN as a function of stellar mass, and host galaxy colour, so a direct comparison to our results is tricky. However, these authors do see a substantial fraction of AGN hosts being non-central galaxies, especially at lower, $M_{\star} < 10^{10.5} M_{\odot}$ stellar mass (their Figure 7). 

\cite{2016ApJ...832..111J} study the AGN population at low redshift and found that Type I and Type II AGN resides in dark matter haloes with similar masses. But the satellite fraction of Type I AGN are smaller than the Type II AGN. They suggest that this has interesting implications in the QSO unified model as they detect environmental differences between Type I and Type II AGN. It will be interesting to see in the future if a similar difference can be observed at high redshift using our model.

\subsection{Host galaxy type for eBOSS QSOs}\label{sec:discussion_lrgs}
In the local  Universe we observe a relationship between SMBH and host-galaxy bulge 
velocity dispersion ($\sigma$) and luminosity \citep{Bell2008, Gultekin2009, Volonteri2010, McConnell2013, KormendyHo2013}; supermassive Black holes are known to be related to their host bulges. 
Thus, it is not surprising that galaxies with generally massive bulges, such 
as LRGs, would be considered QSO host galaxies. 

Returning to \citet{WangLi2019}, they find that AGN in galaxies with blue colours at all masses, or massive red galaxies with $M_{\star}\gtrsim10^{10.5} M_{\odot}$, show almost identical clustering amplitudes at all scales to control galaxies of the same mass, colour, and structural parameters. 

As mentioned above, direct comparison is difficult since our results are reported as a function of DM halo mass, and \citet{WangLi2019} report stellar mass. Moreover, there is definitely not a 1-to-1 mapping of the red/blue galaxy population in \citet{WangLi2019} to the LRG/ELG population we study. That acknowledged, we can look at broad trends. \citet{WangLi2019} find that there is slight preference for AGN to trigger in red satellite galaxies. This is broadly consistent with our findings. \citet{WangLi2019} also find the blue AGN are less likely to be satellite then the general blue galaxy population. This is again consistent with our findings (our Figure 3, middle panel), which shows that if we allow $f_{\rm sat}$ to be a free parameter, then the QSO satellite fraction is higher at lower halo mass, but become less than the blue galaxies at higher halo mass. We find the transition halo mass is around $M_{\rm halo}\sim10^{13} M_{\odot}$.
\citet{WangLi2019} model their results with a simple halo model -- where central fraction of the AGN is the only free parameter -- and place AGN preferentially in the dark matter halo centres, but requiring a mass-dependent central fraction. Their results suggest that the mass assembly history of dark haloes may play an additional role in the AGN activity in low-mass red galaxies.

Interestingly, \citet{Krishnan2020} note that the most important property in determining the AGN clustering signal is the fraction of AGN in passive host galaxies. This is true for our study as well using the inherent model. We note that our results are in contrast with those of \citet{2014ApJ...780..162M}, who studied high signal to noise spectra and inferred a very small red fraction for QSO hosts.

\subsection{Quasars at intermediate redshift are not special, but they are often satellites}\label{sec:discussion_implications}
The results in our paper illuminate several outstanding issues in QSO physics. 
The mean halo mass of QSO has historically been measured to be a few $\times10^{12} M_{\odot}$, 
\citep[e.g.][]{Croom2005, Ross2009, White2012}, essentially corresponding to a small group mass 
\citep[cf. The Local Group having a total mass of $5.27\times10^{12} M_{\odot}$: ][]{LiWhite2008}. 

\citet{Yoon2019} find that, on average, both massive quasars and massive galaxies reside in environments more than $\simeq\,$2 times as dense as those of their less massive counterparts with $\log_{10}(M_{\rm BH}/M_{\odot})<9.0$. However, massive quasars reside in environments about $\sim$2 times less dense than inactive galaxies with $\log_{10}(M_{\rm BH}/M_{\odot})\gtrsim9.4$, and only about one third of massive quasars are found in galaxy clusters, while about two thirds of massive galaxies reside in such clusters. This indicates that massive galaxies are a much better signpost for galaxy clusters than massive quasars.

This is also what we are seeing. We are also finding QSOs to be hosted by galaxies with massive bulges, i.e. the LRGs. But the key thing is that neither the QSOs or the LRGs are necessarily central galaxies. This explains why the QSOs generally have a lower mean clustering amplitude than the massive galaxies, unless they were e.g. radio loud objects \citep{Wake2008, Shen2009}, such as the radio loud giant ellipticals (e.g. M87 or similar). Thus, a key conclusion of our work is that quasars are not special, but they may well often be satellites.

Also at $z \sim 4$, recent studies of the quasar environment find no strong evidence of luminous quasars to reside in dense environments or be associated with proto-clusters (see, e.g., \citet{Uchiyama18}; also \citet{Overzier2016} and references therein). They do find quasars to reside in small size haloes that that are much more in accord with typical average halo masses found at lower redshifts \citep{ef15, White2012} than reported halo mass by \citet{Shen2007}.

 We are also finding that QSOs can inhabit smaller haloes, of the kind that are dominated by star-forming galaxies. Possibly these two classes of object have different triggering mechanisms, but as far as optical luminosity is concerned we can not differentiate the two cases.
 

\section{Conclusions}
\label{sec:conclusion}
In this paper we have used the final SDSS eBOSS DR16 spectroscopic dataset for luminous red galaxies, emission line galaxies and QSOs to make multi-tracer clustering measurements. The motivation is to {\it (i)} investigate how the QSO population samples the galaxy population; and {\it (ii)} to understand how the QSO host dark matter haloes sample the underlying dark matter halo distribution. 

Our main conclusions are: 
\begin{itemize} 
\item The probability that a galaxy has quasar activity is independent of dark matter halo mass.
\item QSOs host galaxies have a large satellite fraction, probably due to their low number density (and this is possible, even without a large one-halo term). 
\item We infer the halo mass distribution of QSOs to be very broad, independent of assumptions in modelling about the parent population. 
\item All QSOs can {\it not\/} be in LRG host galaxies (at more than the 3$\sigma$ level). 
\item Likewise, all QSOs cannot correspond only to ELG host galaxies (at the $\sim$2$\sigma$ level). 
\item Given that the spline model works and that the parameter $f_{\rm on}$ is flat, the error function is generally a good model to describe $N_{\rm QSO}(M_{\rm halo})$.
\end{itemize} 

The discussion of how environmental influences or assembly bias affects the QSO population is left for future studies. 
In the broadest sense, this self-sufficient study, in which the internally observed and measured correlation functions constrain the characteristics of the native halo catalogue, provides a much more self-consistent picture to the nuanced world of quasar occupation: {\bf (i)} This likelihood-driven study grants a fair chance to multiple characteristic dependencies to arise from the measurement. The inferred indifference of the satellite fraction to the host halo mass in this picture, distinguishes itself from the previous studies and reported discrepancies where the partially outsourced measurement, halo model or adapted halo mass distribution, invite a host of added assumptions and justifications. See, e.g., \citet{Eftekharzadeh2019} and \citet{KayoOguri2012} for similarly adapted functional forms for the halo profile and halo mass distribution model, and yet significantly different satellite fraction that was attributed to a plausible luminosity dependency of the small scale clustering as opposed to lack thereof on large scales. {\bf(ii)} This study highlights the notion that sampling quasars that belong to a group or are otherwise members of a close-pair system (a.k.a. `binary'), leads to diverging conclusions on satellite occupation (see $\sec$ \ref{sec:discussion_fsat}). This could be viewed as further evidence for hierarchical growth studies that have found earlier formation times for close halo pairs compared to their distant counterparts \citep{Sheth1999,Harker2006}.

{\bf (iii)} Halo mass measurement for quasars in early eBOSS data inferred two plausible scenarios for the relatively constant characteristic halo mass between $z\sim 1$ and $z \sim 2$ (Figure 10 in \citet{Laurent2017}) to be either due to massive haloes dominating the average or less luminous quasars inhabiting a wide range of halo masses and therefore putting the moderately luminous quasars in a different evolutionary path than the $\sim 4000$ highly luminous quasars sampled by \citet{Shen2007} with a dramatically higher average halo mass. The inferred broad halo mass range in this study provides an elaborate case for the latter scenario using quasars in the same parent sample and luminosity class.

As the next generation surveys including 
eROSITA \citep[][]{eROSITA}, 
SKA\footnote{\url{https://www.skatelescope.org/the-ska-project/}}, 
ESA {\it Euclid}, 
the Dark Energy Spectroscopic Instrument \citep[DESI; ][]{2016arXiv161100036D}, 
the Prime Focus Spectrograph \citep[PFS; ][]{2014PASJ...66R...1T} 
now come online, we will be able to test our findings and fully investigate the host galaxy population of luminous AGN across cosmic history. 

\section*{Acknowledgments}
We thank Andy Lawrence for useful discussions. SA and JAP are supported by the European Research Council through the COSFORM Research Grant (\#670193).  
NPR acknowledges support from the STFC and the Ernest Rutherford Fellowship scheme. 

We thank Horst Meyerdierks and Eric Tittley for their support with Stacpolly and Cuillin cluster where all of the computing for this project is performed. We thank
the Multi Dark Patchy Team for making their simulations publicly available. This research has made use of NASA's Astrophysics Data System.
The CosmoCalculator was used \citep{Wright2006}. 

Funding for the Sloan Digital Sky Survey IV has been provided by the Alfred P. Sloan Foundation, the U.S. Department of Energy Office of Science, and the Participating Institutions. SDSS-IV acknowledges
support and resources from the Center for High-Performance Computing at
the University of Utah. The SDSS web site is www.sdss.org.

SDSS-IV is managed by the Astrophysical Research Consortium for the 
Participating Institutions of the SDSS Collaboration including the 
Brazilian Participation Group, the Carnegie Institution for Science, 
Carnegie Mellon University, the Chilean Participation Group, the French Participation Group, Harvard-Smithsonian Center for Astrophysics, 
Instituto de Astrof\'isica de Canarias, The Johns Hopkins University, Kavli Institute for the Physics and Mathematics of the Universe (IPMU) / 
University of Tokyo, the Korean Participation Group, Lawrence Berkeley National Laboratory, 
Leibniz Institut f\"ur Astrophysik Potsdam (AIP),  
Max-Planck-Institut f\"ur Astronomie (MPIA Heidelberg), 
Max-Planck-Institut f\"ur Astrophysik (MPA Garching), 
Max-Planck-Institut f\"ur Extraterrestrische Physik (MPE), 
National Astronomical Observatories of China, New Mexico State University, 
New York University, University of Notre Dame, 
Observat\'ario Nacional / MCTI, The Ohio State University, 
Pennsylvania State University, Shanghai Astronomical Observatory, 
United Kingdom Participation Group,
Universidad Nacional Aut\'onoma de M\'exico, University of Arizona, 
University of Colorado Boulder, University of Oxford, University of Portsmouth, 
University of Utah, University of Virginia, University of Washington, University of Wisconsin, 
Vanderbilt University, and Yale University. 

The CosmoSim database used in this paper is a service by the Leibniz-Institute for Astrophysics Potsdam (AIP).
The MultiDark database was developed in cooperation with the Spanish MultiDark Consolider Project CSD2009-00064.

The authors gratefully acknowledge the Gauss Centre for Supercomputing e.V. (\url{www.gauss-centre.eu}) and the Partnership for Advanced Supercomputing in Europe (PRACE, \url{www.prace-ri.eu}) for funding the MultiDark simulation project by providing computing time on the GCS Supercomputer SuperMUC at Leibniz Supercomputing Centre (LRZ: \url{www.lrz.de}).

The Hyper Suprime-Cam (HSC) collaboration includes the astronomical communities of Japan and Taiwan, and Princeton University. The HSC instrumentation and software were developed by the National Astronomical Observatory of Japan (NAOJ), the Kavli Institute for the Physics and Mathematics of the Universe (Kavli IPMU), the University of Tokyo, the High Energy Accelerator Research Organization (KEK), the Academia Sinica Institute for Astronomy and Astrophysics in Taiwan (ASIAA), and Princeton University. Funding was contributed by the FIRST program from the Japanese Cabinet Office, the Ministry of Education, Culture, Sports, Science and Technology (MEXT), the Japan Society for the Promotion of Science (JSPS), Japan Science and Technology Agency (JST), the Toray Science Foundation, NAOJ, Kavli IPMU, KEK, ASIAA, and Princeton University. 

This paper makes use of software developed for the Large Synoptic Survey Telescope. We thank the LSST Project for making their code available as free software at  \url{http://dm.lsst.org}.

This paper is based [in part] on data collected at the Subaru Telescope and retrieved from the HSC data archive system, which is operated by Subaru Telescope and Astronomy Data Center (ADC) at National Astronomical Observatory of Japan. Data analysis was in part carried out with the cooperation of Center for Computational Astrophysics (CfCA), National Astronomical Observatory of Japan.

\section{Data Availability}
All of the observational datasets used in this paper are available through the SDSS website \url{https://data.sdss.org/sas/dr16/eboss/}. The codes used in this analysis along with instructions are available on \url{https://www.roe.ac.uk/~salam/GQOD/}.




\bibliographystyle{mnras}
\bibliography{tester_mnras,Master_Shadab}


\appendix



\section{QSO 50\% Bright sample}
\begin{figure*}
    \centering
    \includegraphics[width=1.0\textwidth]{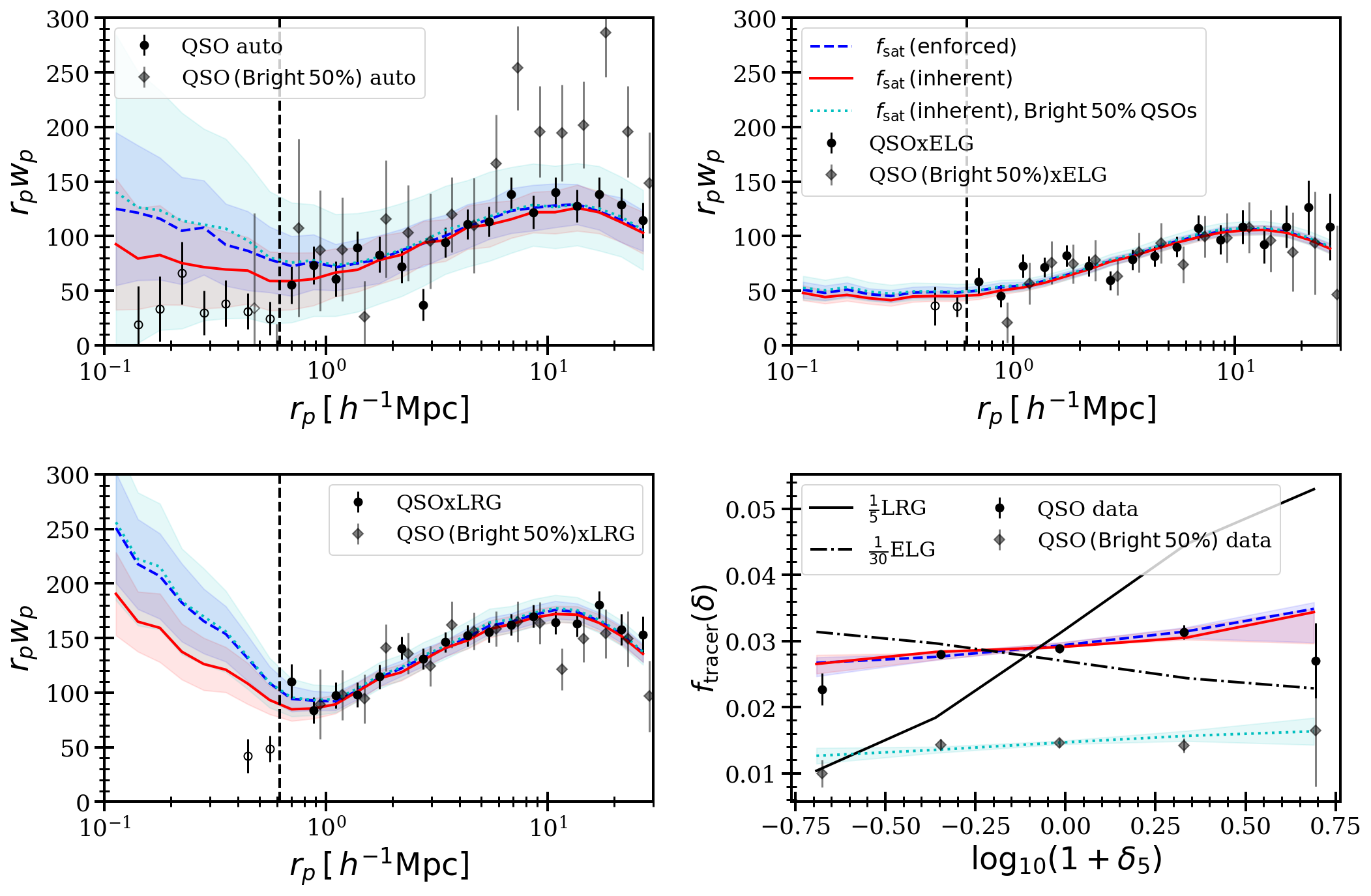}
    \caption{ Same as Figure \ref{fig:wp-auto-cross} but including QSO Bright 50\% sample. The Diamond points in each panel shows the measurements from the Bright QSO sample where as dotted cyan line with shaded region shows best fit model to QSO Bright sample along with 1$\sigma$ spread.
     }
    \label{fig:wp-auto-cross-wQSOLum}
\end{figure*}

Figure ~\ref{fig:wp-auto-cross-wQSOLum} shows a version of data and best fit model including QSO Bright sample.


\bsp	
\label{lastpage}
\end{document}